**Searching for small-world and scale-free behaviour in long-term historical data of a real-world power grid**

Bálint Hartmann*, Viktória Sugár

Centre for Energy Research, Environmental Physics Department, KFKI Campus, Konkoly-Thege Miklós út 29-33., 1121 Budapest, Hungary

* Corresponding author. Tel.: +36 20 4825310. E-mail address: hartmann.balint@energia.mta.hu, ORCID: 0000-0001-5271-2681

**Abstract**

Since the introduction of small-world and scale-free properties, there is an ongoing discussion on how certain real-world networks fit into these network science categories. While the electrical power grid was among the most discussed examples of these real-word networks, published results are controversial, and studies usually fail to take the aspects of network evolution into consideration. Consequently, while there is a broad agreement that power grids are small-world networks and might show scale-free behaviour; although very few attempts have been made to find how these characteristics of the network are related to grid infrastructure development or other underlying phenomena. In this paper the authors use the 70-year-long historical dataset(1949–2019) of the Hungarian power grid to perform complex network analysis, which is the first attempt to evaluate small-world and scale-free properties on long-term real-world data. The results of the analysis suggest that power grids show small-world behaviour only after the introduction of multiple voltage levels. It is also demonstrated that the node distribution of the examined power grid does not show scale-free behaviour and that the scaling is stabilised around certain values after the initial phase of grid evolution.

**Introduction**

It was a little more than twenty years ago when two papers gave impetus to the field of network sciences. In their paper [1] Watts and Strogatz presented the concept of small-world networks, describing systems that are highly clustered but have small characteristic path lengths, thus showing similarity in certain aspects to lattices and random graphs as well. A year later Barabási and Albert reported [2] the discovery of a high degree of self-organization in large complex networks based on the nature of the interaction between vertices (nodes), an attribute to become known as scale-free behaviour. Both papers demonstrated their concepts on real-world networks, among which a common choice was the electrical power grid of the Western United States (modelled as nodes being generators, transformers and substations and the edges being the power lines between them).

These findings have received outstanding attention from the scientific community, which has led to a number of studies disputing their initial findings and statements, especially with respect to scale-free behaviour. Amaral et al. [3] noted that the degree distribution of power grids is better fitted by an exponential distribution than with a power-law, especially in the case of lower degree nodes. Aging and the limited capacity of nodes were named as potential causes of this difference. This finding was confirmed by Albert et al. [4]. The small-world behaviour was in the focus of Cloteaux's work [5], which concluded that a generalization that all power grids show small-world characteristics cannot be made. He assumed that the structure of the power grid is reminiscent of certain complex networks because of the underlying population distributions.

Despite these early critiques, a large number of papers were published on the topic in the first decade of the new millennium. Authors have examined the small-world behaviour for various power grids, including the Western US [6] [7] [8], North America [9] [10], China [9] [11] **Error! Reference source not found.** [13], Scandinavia [7], Europe [14] and The Netherlands [15]. Different distributions were fitted to the cumulative probability distributions of node degrees, including exponential [4] [9] [13] [14] [16] [17] [18], power-law [19] and mixed models [8] [15] [20] [21], leading to somewhat controversial results.

While it seems that the focus of publications has shifted to new topics, the debate is far from over, as highlighted by the comment of Holme [22]. Influential papers include the ones published by Clauset, Shalizi and Newman [23] and Broido and Clauset [24] claiming that technological networks exhibit very weak or no evidence of scale-free structure. This work was recently re-evaluated by Artico et al. [25], leading to opposite results, classifying almost two-thirds of the examined networks as scale-free. While being very comprehensive, the latter two studies only marginally touched upon power grids (as of today, the Colorado Index of Complex Networks database includes only 4 power grid topologies of the total 5410 entries), thus leave the questions of small-world and scale-free behaviour open in that field.

The number of papers specialized on this topic is also very low. Buzna et al. [26] analysed the 40-year-long evolution of the French 400 kV transmission network, which was characterized by a slow phase, followed by an intensive growth and a saturation. Their most important findings were that a small-world property was only seen in the saturation phase of the process, and that the clustering coefficient of the network has started to decrease after 1996. Buzna et al. did not consider the importance of multiple voltage levels in small-world behaviour, which was addressed by Espejo et al. [27] by examining 400 kV and 220 kV networks of 15 countries. While concluding that all selected networks can be considered as small-world, this is only true if voltage levels are considered jointly. Their work did not consider the evolutionary aspect of power grid development either.

Analytical models of power grid evolution were presented by Deka et al. [28], showing that the node degree distribution of the generated synthetic networks was a weighted sum of shifted exponentials, similar to what is observed in the case of many European and American power grids.

A small-world model was used to simulate the 50-year-long growth and evolution of a power grid consisting of multiple voltage levels by Mei et al.**Error! Reference source not found.**. It was shown that the used evolution model did not lead to consistent characteristics of node degree distribution. While certain snapshots showed scale-free behaviour, no generalization could be made. It is also worth noting that exponential distributions were not a good fit for the node degree distribution either.

To conclude, results of available literature do not clarify the question whether power grids can be handled as small-world and/or scale-free networks, whether they display such behaviour on the long-run or only for certain periods, and whether this is the result of an evolutionary process related to grid infrastructure development or possibly different underlying phenomena. The present paper aims to contribute to all three fields, by examining the 70-year-long historical development of the Hungarian power grid (1949–2019), including all voltage levels (120 kV, 220 kV, 400 kV and 750 kV) that have been constructed for transmission networks in different periods of time. Using the historical dataset, the authors have created graph representations for each year and examined various network properties, while relating the change of those properties to major system changes, which may uncover underlying causes. To the authors' knowledge, no such long-term evaluation of the topic has been published yet.

**Results and discussion**

The results of the long-term network analysis are shown in **Fig. 1.** In general, it can be observed that most properties only exhibit variations in the first two decades of the evaluated period, while from 1970 on, the majority of the values can be handled as constants, as discussed in the following.

The number of nodes ($N$) and edges ($E$) show similar increasing trends, $E$ being significantly higher after 20 years. The average node degree of the network ($\langle k \rangle$) varies between 1.8 and 2.61, the values marking the first and the last year of the evaluated period. The value first exceeds 2.5 in 1969 and remains practically constant for the next 50 years. The diameter of the network ($d$) is the lowest in 1949 (5) and the highest in 1970 (18). Its value varies between 14 and 15 since 1978. The modularity quotient ($Q$) of the network is in the same range since 1971 (between 0.453 and 0.469).

A different behaviour is shown by the average path length ($L$) and the clustering coefficient ($C$) of the network. $L$ shows a very fast increase in the first two decades, which later transforms to a less steep but still steady trend. It is also important that the $L$ of the actual network is bigger than the $L_r$ of the respective random network, which is necessary to be characterised as a small-world network.

The biggest variations are shown by the clustering coefficient $C$. In the first decade of the evolution of the network, $C$ differs from zero only once. In the next two decades, four important increases can be seen, between 1958–1961, 1965–1969, 1977–1978 and 1985–1990. These increases can be connected to well-identifiable network development activities. In the first period, the Tiszapalkonya–Szolnok line and the construction of the Oroszlány power plant with three connecting lines modified the topology of the Eastern and Western part of the system significantly, respectively. The second period marks the construction of the backbone of the 220 kV system: 7 lines were commissioned (Dunamenti–Zugló, Dunamenti–Soroksár, Sajószöged–Szolnok, Zugló–Göd, Sajószöged–Detk, Detk–Zugló and Detk–Szolnok). This development, on the one hand, connected distant parts of the national network, and on the other hand, formed a meshed topology. During the third period, the main elements of the 400 kV system were constructed (9 lines: Tisza Power Plant–Sajószöged, Albertfalva–Dunamenti, Sajószöged–Göd, Sajószöged–Felsőzsolca, Dunamenti–Martonvásár, Litér–Martonvásár, Martonvásár–Toponár and Győr–Litér). The last period marked the construction of the Paks Nuclear Power Plant and the commissioning of three 400 kV lines (Litér–Toponár, Paks–Litér, Albertirsa–Békéscsaba), which have again created a meshed formation in the network. These new formations are shown in **Fig 2.**

It is evident that these periods also influenced the small-world coefficient ($\sigma$) of the network. The value of $\sigma$ becomes bigger than unity for the first time in 1968 and reaches its final range in 1990. The relationship of $\sigma$ and the number of 220 and 400 kV lines within the borders shows high correlation (0.91) with the time-series data (**Fig. 3.**). Based on these results, the authors conclude that the Hungarian power grid has shown small-world properties only after the introduction of multiple voltage levels into the transmission network. The technical description of such network development (connecting distant points of the network with higher voltage levels to decrease transmission losses) also resembles the method of creating small-world networks. This conclusion is similar to the one drawn in [27] as presented in the introduction, emphasizing the importance of multiple voltage levels.

Power-law and exponential fits to the cumulative node degree distribution for 10-year snapshots are shown in **Fig. 4.** and **Fig. 5.**, respectively. Both figures show that parameters of the fitted distribution show little variance from 1979, despite that the size of the network has increased from 220 nodes and 281 edges to 385 nodes and 504 edges. It can be clearly seen that power-law fit performs poorly for high-degree nodes, while in the case of the exponential fit, such a major issue is not seen. Numerical values of the fits were compared to the ones reported in the literature and are in the same range for exponential fits but show a large difference for power-law fits. Based on these results, the authors conclude that the node distribution of the examined long-term model does not show scale-free behaviour and that the scaling of the network varies in a relatively small range, which may characterise the specific network evolution process examined.

It is also worth comparing the presented results to the only paper discussing a long-term grid evolution [12], since the parameters of the grid simulated by Mei et al. shows remarkable similarities to the Hungarian power grid. Selected values after 50 years of development are compared in **Table 1**. The number of nodes and edges, the installed capacity of power plants, the clustering coefficient and the average node degree are very close to each other. A difference is seen in the distribution of lines with various voltage levels; this is mainly due to the slightly different physical area of the grids and the waiting time before the introduction of new voltage levels. Mei et al. allow 220 and 500 kV lines after 19 and 39 years, respectively; while in the Hungarian grid it actually took 13 and 29 years, respectively. As for the small-world properties of the networks, the simulated grid was considered to show such properties after approximately 20 years, which was the same in the case of the historical data of the Hungarian grid.

**Conclusions**

Long-term historical data of the Hungarian power grid was examined using tools of complex network analysis to search for small-world and scale-free behaviour. It was observed that most properties stabilized at practically constant values after the initial phase of grid evolution. This initial phase took approximately 20 years and was closed by the introduction and deployment of the 220 kV voltage level, which connected distant nodes of the network, and formed a meshed topology. Four periods of grid development were identified, during which the clustering coefficient (and thus the small-world coefficient) of the network has increased significantly. All of these periods were related to the introduction of new voltage levels and the creation of meshed/looped topological formations, which is atypical in single voltage level subnetworks of the power grid. The authors have concluded that power grids show small-world behaviour only if they consist of multiple voltage levels. Power-law and exponential fits to cumulative node degree distributions have shown that power-law fits perform poorly for nodes with high connectivity, thus the use of exponential fits should be preferred.

**Methods and data**

*Network data*

The authors have assembled the network data using various sources, including hand-written notes, anniversary books [29] [30] [31] [32] [33], statistical publications, maps and personal consultation. Since none of the sources were consistent, certain pre-processing and standardisation had to be made. In the database, a new node was created when a substation was first constructed, regardless of the installed switchgear and the type of the busbar. A new edge was created when a power line was put into operation. Double systems are handled as single connections. Infrastructural elements were removed from the database in the year of decommissioning. The final database spans over 70 years and includes 400 nodes and 774 edges.

Synchronous operation of the Hungarian power grid began in 1949, when 6 power plants and 11 substations were controlled by the National Load Dispatch Centre (Országos Villamos Teherelosztó – OVT in Hungarian. In the first half of the 1950s, primarily as a result of the dynamic industrial developments, power plant construction could not keep pace with the electricity demand increase; but the extension of the network made it possible to connect existing small power plants into the co-operation. The first international connection was put into operation in 1952 (Kisigmánd–Nové Zámky, Slovakia 120 kV). The 220 kV voltage level was first introduced in 1960 (Zugló–Bistričany, Slovakia), which year marked the beginning of the development of the 220 kV transmission grid, with this voltage level becoming dominant over the next decades. Connections were formed to the east from Sajószöged and Zugló substations, and then to the west from Győr. By the mid-1970s, more than 1,000 km of 220 kV transmission lines were in operation. The second half of the 1970s witnessed a spectacular development of the 400 kV network, and finally the 750 kV voltage level was introduced in 1978. The next period was aimed at creating loops to mesh the 400 kV network both domestically and internationally. The latest change to be mentioned came in 1992, when the 120 kV sub-transmission network became the property of utility companies that were turned into independent corporations, while the 220, 400 and 750 kV network became the property of the national transmission system operator. From this time on, the so-called ($n$-1) security criterion had to be solely satisfied by the transmission grid, which has determined network development ever since. As of 2019, the Hungarian power grid consisted of 266 km of 750 kV lines, 2297 km of 400 kV lines, 1099 km of 220 kV lines and 6536 km of 132 kV lines, altogether 385 nodes and 504 edges. 10-year snapshots of the grid evolution are shown in **Fig. 6.**

*Graph properties*

Using the presented dataset, graph representations were made for each year, with the nodes being generators, transformers and substations and the edges being transmission lines. For the graphs, the authors have calculated the node degree distribution and average node degree, the diameter, the modularity metric, the average path length, the clustering coefficient and the small-world metric. Of those properties, node degree distribution and diameter are well-known, thus the authors restrict themselves to discussing the remaining ones.

The modularity quotient, *Q* is defined as:

$$Q = \frac{1}{N\langle k \rangle} \sum_{ij} \left( A_{ij} - \frac{k_i k_j}{N\langle k \rangle} \right) \delta(g_i, g_j) \quad (1)$$

where $A_{ij}$ is the adjacency matrix and $\delta(i,j)$ is the Kronecker-delta function.

The average path length, *L* is defined as:

$$L = \frac{1}{N(N-1)} \sum_{j \neq i} d(i,j) \quad (2)$$

where $d(i,j)$ is the graph distance between nodes *i* and *j*. For random networks, *L* is obtained by the following formula [34]:

$$L_r = \frac{ln(N) - 0.5772}{ln\langle k \rangle} + 0.5 \quad (3)$$

As defined in [1], the clustering coefficient *C* is defined as:

$$C = \frac{1}{N} \sum_i \frac{2E_i}{k_i(k_i - 1)} \quad (4)$$

where *E* is the number of edges between the neighbours of *i*. The clustering coefficient for random networks is determined as:

$$C_r = \frac{\langle k \rangle}{N} \quad (5)$$

Using Eqs. (2–5) the small-word coefficient σ can be calculated as follows [34]:

$$\sigma = \frac{C/C_r}{L/L_r} \quad (6)$$

A network can be considered small-world if $C \gg C_r$ and $L \geq L_r$, i.e. if σ is larger than unity.

To examine whether the networks display scale-free behaviour, the authors have fitted both exponential and power-law distributions to the node degree distribution. As it was noted in the Introduction of this paper, consensus currently is that exponential fits are correct, but as power grids are still widely considered to show scale-free properties, this second option was examined as well. Fits were prepared using the *'fit'* function of MATLAB R2019a.

**Acknowledgements**

The authors would like to thank the following people for their contribution to assembling the historical grid database: Attila Dervarics, Zoltán Feleki, József Hiezl, János Nemes, Imre Orlay, János Rejtő.

**Author contributions**

Bálint Hartmann conceived and designed the analysis, performed the analysis and wrote the paper. Bálint Hartmann and Viktória Sugár collected the data and contributed analysis tools. Viktória Sugár prepared topological maps.

**Competing interests**

The authors declare no competing interest.

**Figure legends**

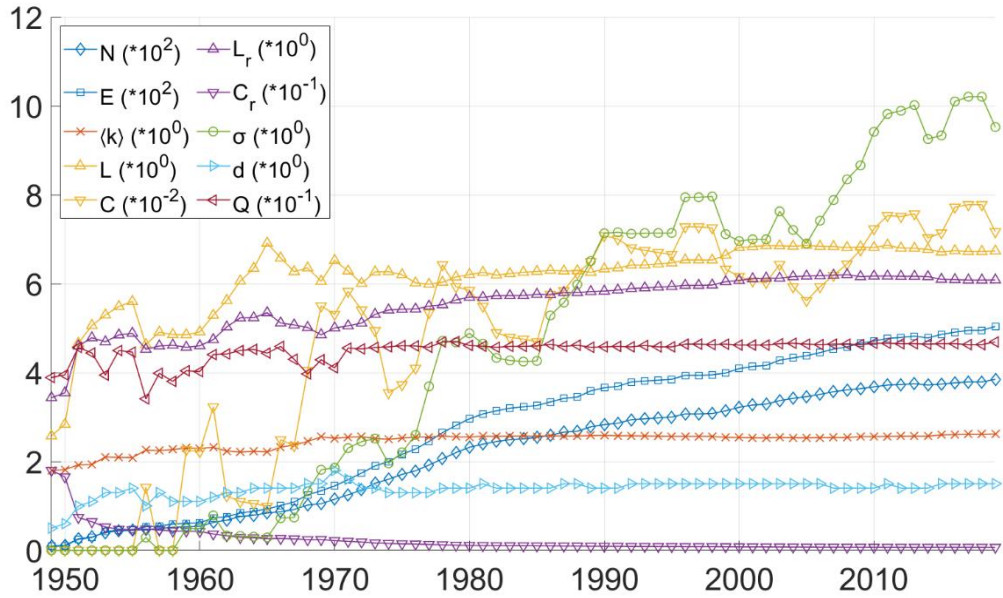

**Figure 1.** Results of the long-term complex network analysis. *N* (nodes), *E* (edges), $\langle k \rangle$ (average node degree), *L* (average path length), *C* (clustering coefficient), $L_r$ (average path length of random network), $C_r$ (clustering coefficient of random network), $\sigma$ (small-world coefficient), *d* (diameter), *Q* (modularity quotient)

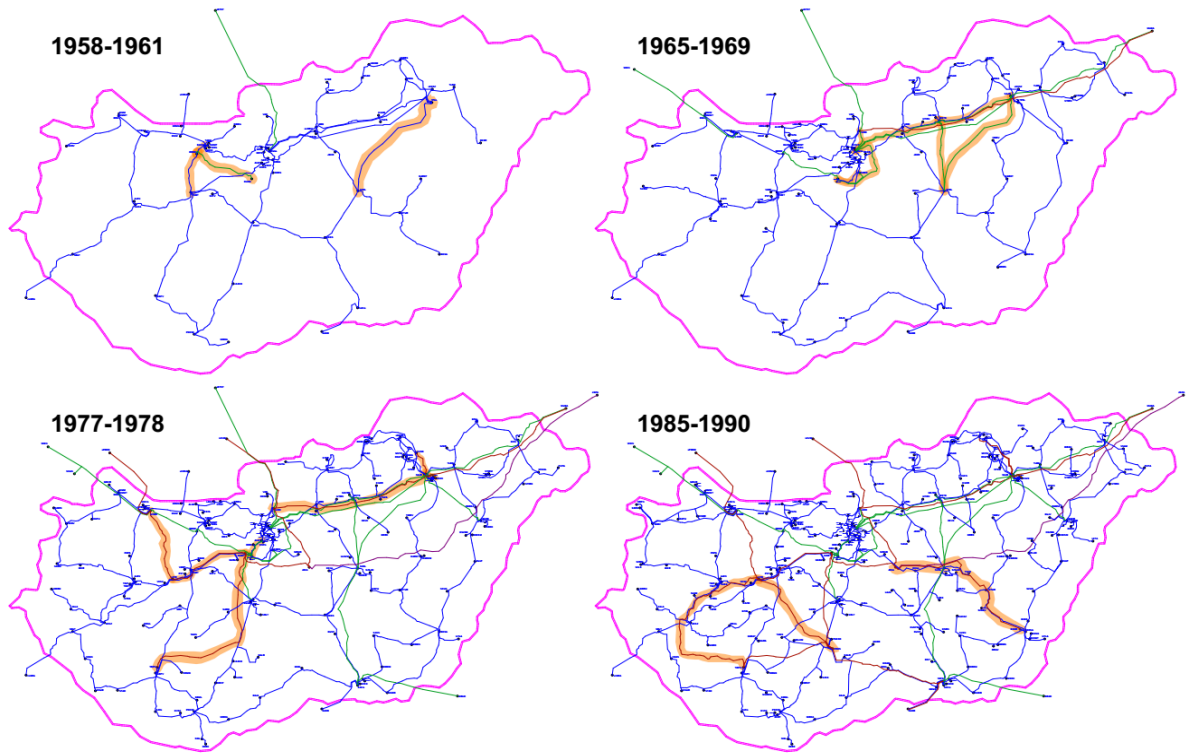

**Figure 2.** Four periods of network development activities, which have significantly increased the clustering coefficient of the network. A vast majority of the newly commissioned connections were 220 and 400 kV lines, creating a meshed topology with the underlying 120 kV network.

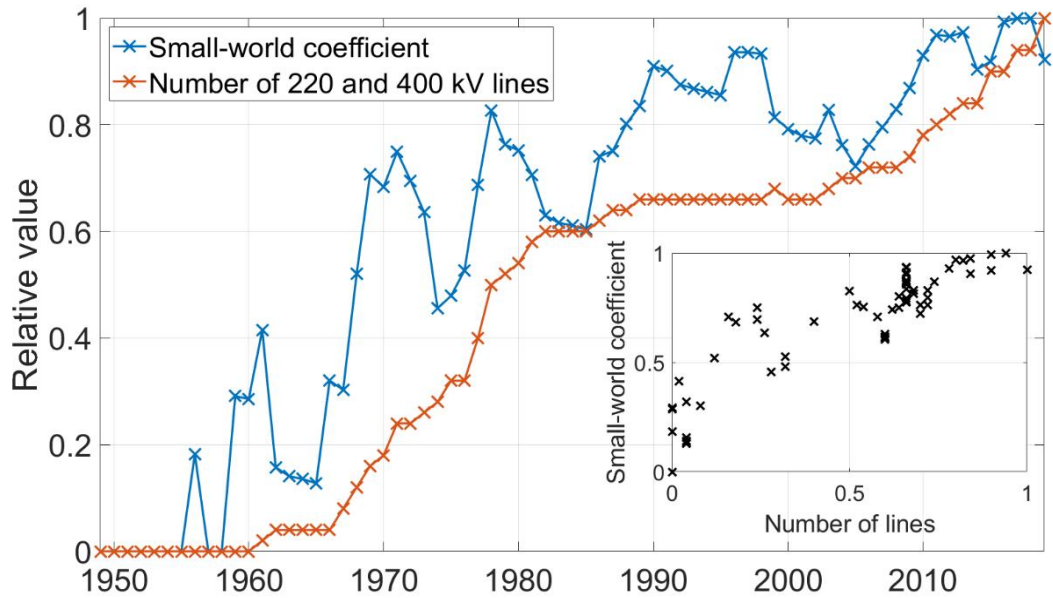

**Figure 3.** Relationship between the number of domestic 220 and 400 kV lines and the small-world coefficient. The main plot shows relative values compared to the maximum achieved during the 70-year-long period. The inset shows the nature of the correlation.

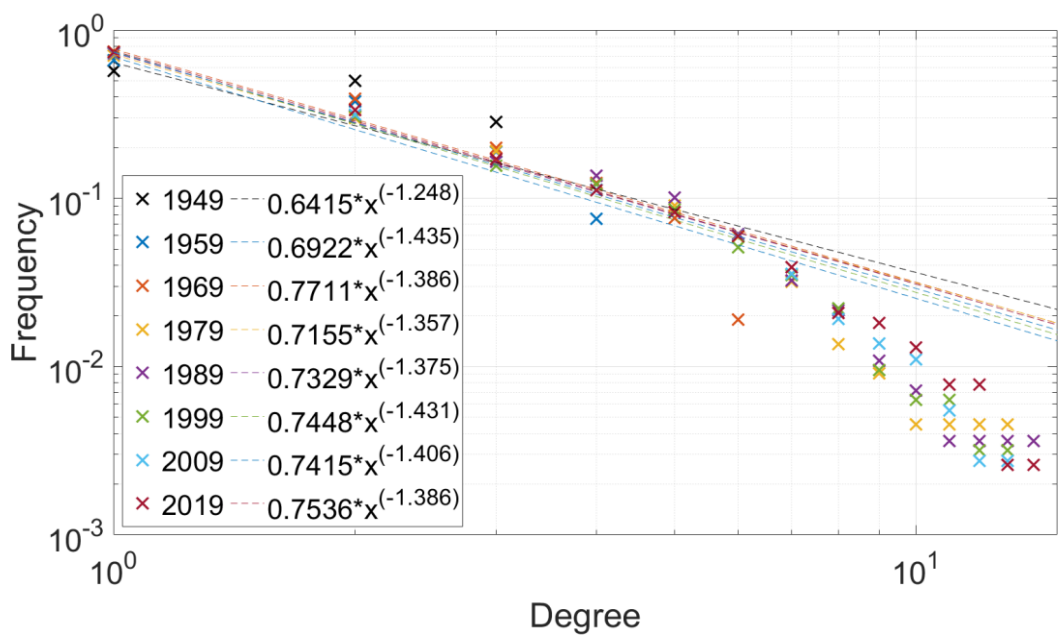

**Figure 4.** Cumulative node degree distribution and power-law fits

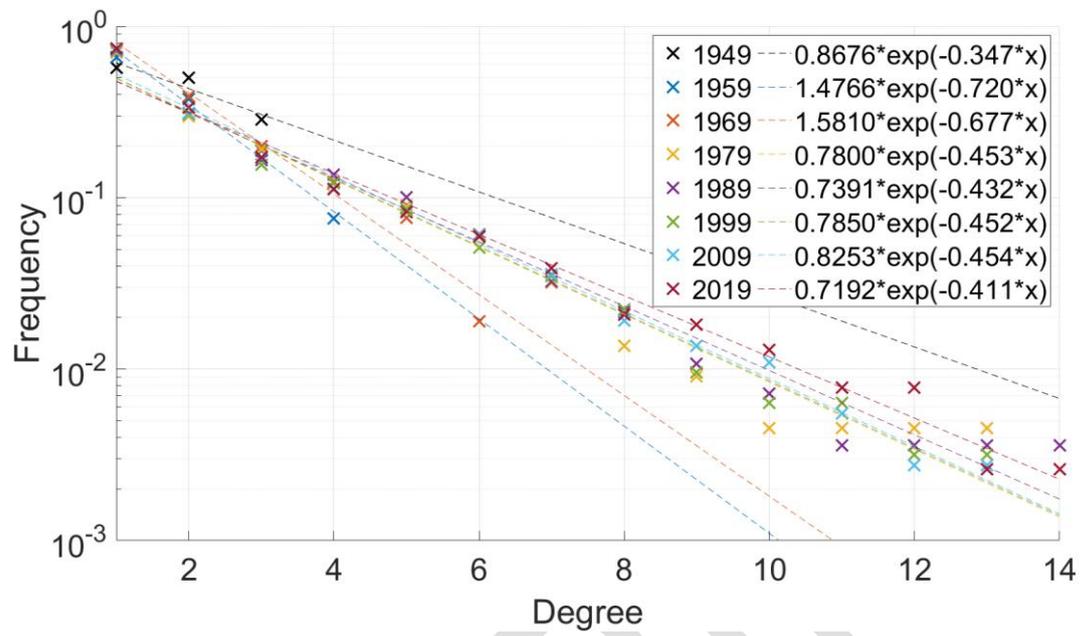

**Figure 5.** Cumulative node degree distribution and exponential fits

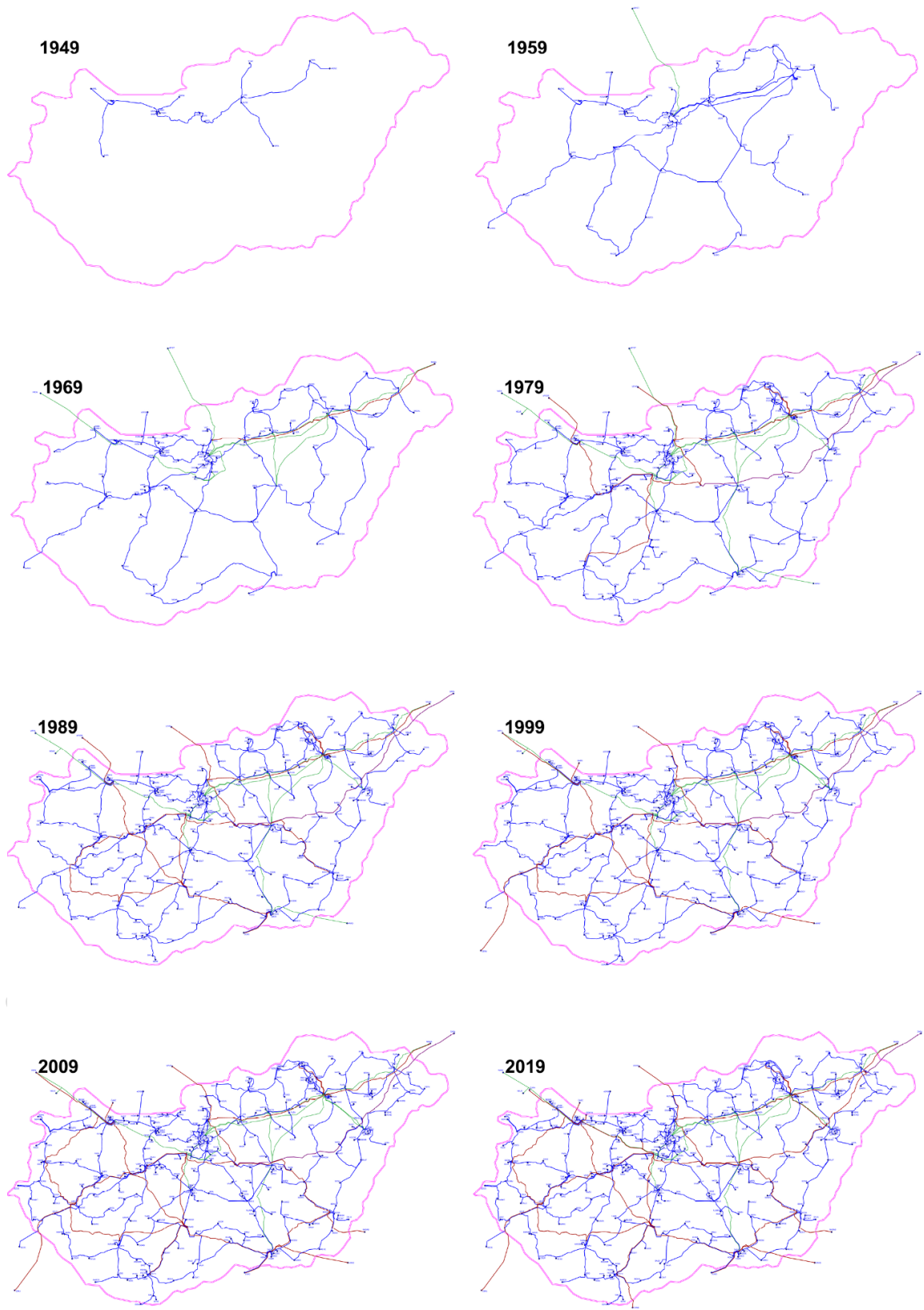

**Figure 6.** The evolution of the Hungarian power grid, snapshots taken every 10 year. Blue lines represent 120 kV, green lines represent 220 kV, red lines represent 400 kV and the purple lines represent 750 kV power lines.

**Tables**

**Table 1.** Comparison of selected parameters of the grid simulated by Mei et al., and the Hungarian power grid

|  |  | Simulated grid [12] | Hungarian grid in 1999 |
|---|---|---|---|
| Nodes |  | 300 | 314 |
| Edges |  | 331 | 399 |
| Installed capacity [MW] |  | 6680 | 5742 |
| Line length [km] | 120 kV | 2799 | 6952 |
|  | 220 kV | 2926 | 1194 |
|  | 400/500 kV | 1766 | 1733 |
|  | Total | 7491 | 9879 |
| C |  | 0.074 | 0.063 |
| L |  | 9.82 | 6.64 |
| $\langle k \rangle$ |  | 2.2 | 2.54 |
| $\sigma$ |  | 7.48 | 7.11 |